\documentclass[
 reprint, 
 amsmath, 
 aps 
]{revtex4-2}

\usepackage{hyperref} 
\usepackage{graphicx} 
\usepackage{orcidlink} 
\usepackage{subfig} 
\usepackage{array} 
\usepackage{enumitem} 
\usepackage[ruled, linesnumbered]{algorithm2e} 

\hypersetup{ 
    colorlinks,
    linkcolor=blue,
    citecolor=blue,
    urlcolor=blue
}

\newcolumntype{C}[1]{>{\centering\arraybackslash}p{#1}}

\setlist{itemsep=-5pt, topsep=5pt}

\begin{document}

\title{Decomposition of Sparse Amplitude Permutation Gates with Application to Preparation of Sparse Clustered Quantum States}

\author{Igor Gaidai\orcidlink{0000-0002-3950-3356}$^1$}
\email{igaidai@utk.edu}

\author{Rebekah Herrman\orcidlink{0000-0001-6944-4206}$^1$}
\email{rherrma2@utk.edu}

\affiliation{$^1$Department of Industrial and Systems Engineering, University of Tennessee Knoxville, USA}

\begin{abstract}

In this work we consider a novel heuristic decomposition algorithm for $n$-qubit gates that implement specified amplitude permutations on sparse states with $m$ non-zero amplitudes.
These gates can be useful as an algorithmic primitive for higher-order algorithms.
We demonstrate this by showing how it can be used as a building block for a novel sparse state preparation algorithm, Cluster Swaps, which is able to significantly reduce CX gate count compared to alternative methods of state preparation considered in this paper when the target states are clustered, i.e. such that there are many pairs of non-zero amplitude basis states whose Hamming distance is 1.
Cluster Swaps can be useful for amplitude encoding of sparse data vectors in quantum machine learning applications.



\end{abstract}

\maketitle

\section{Introduction}
\label{sec:intro}
\begin{figure}
    \centering
    
    \subfloat[Amplitude permutation pattern induced by swapping qubits 2 and 3.]{\includegraphics[width=0.2\textwidth]{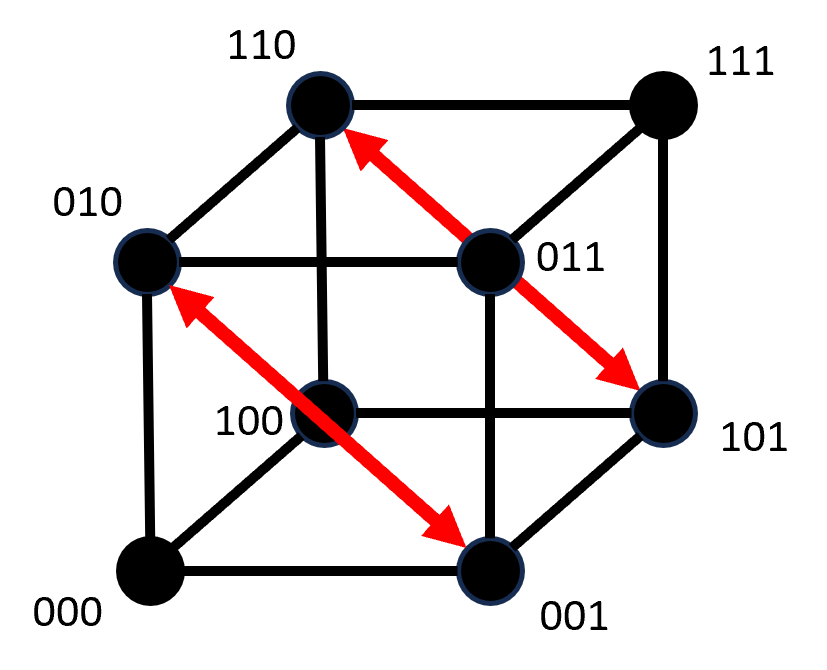}}
    \hfill
    \subfloat[Arbitrary permutation.]{\includegraphics[width=0.2\textwidth]{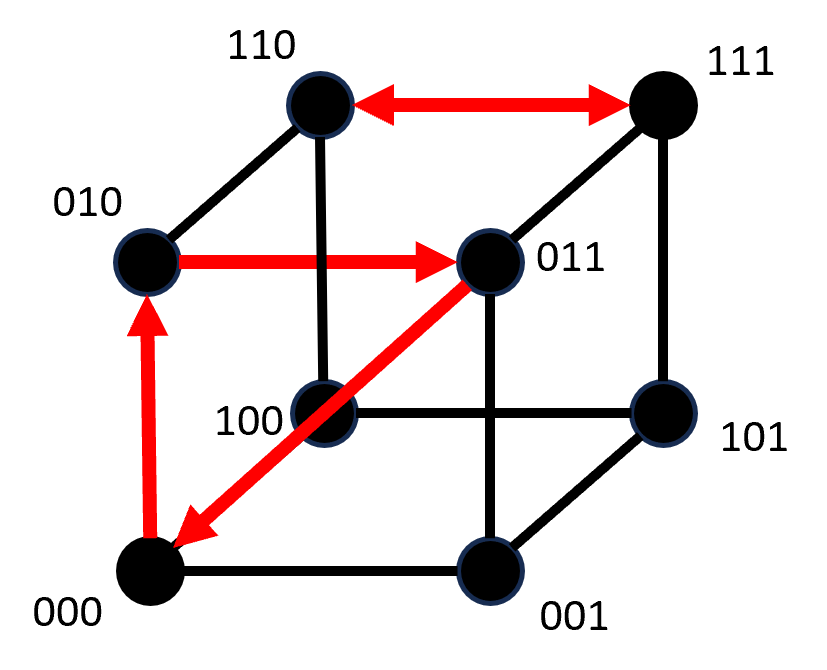}}
    
    \caption{Examples of amplitude permutations for 3 qubits.}
    \label{fig:permutation_examples}
\end{figure}

When designing new quantum algorithms, it is often useful to abstract from the basic gates and think in terms of higher-order algorithmic primitives. 
One such primitive is the operation of amplitude permutation (also known as basis state permutation). 
A special case of this permutation is the permutation of qubits, which is commonly used during compilation of logical quantum circuits with all-to-all connectivity into lower-level circuits that take into account limited hardware connectivity and add necessary swaps to enable interaction between qubits that cannot interact directly \cite{o2019generalized, liu2024realization, childs2019circuit, liu2023tackling, de2019finding}. 
Qubit permutation corresponds to specific regular patterns when considered as an amplitude permutation (Figure~\ref{fig:permutation_examples}a).

More generally, one can consider the operation of arbitrary amplitude permutation that does not conform to qubit-permutation patterns (Figure~\ref{fig:permutation_examples}b). 
Such arbitrary permutations have multiple existing applications, e.g. in cryptography \cite{kuang2022quantum, atmoko2024encryption}, algorithms \cite{vandersypen2000experimental,yalccinkaya2017optimization, metwalli2024testing}, or mixer design \cite{hadfield2019quantum, palackal2023graph, tsvelikhovskiy2024equivariant} for QAOA-like methods \cite{farhi2014quantum, bako2022prog, wilkie2024quantum, li2024quantum, maciejewski2024design, ng2024analytical, lyngfelt2025symmetry, katial2025instance, herrman2022multi, gaidai2024performance}.

An arbitrary amplitude permutation operation may in general require an $n$-qubit gate, which needs to be decomposed into a product of smaller hardware-supported gates before it can be used on a quantum computer \cite{mottonen2006decompositions, rakyta2022efficient, tomesh2024quantum, rosa2025optimizing, liu2023qcontext, vale2023decomposition}.
Prior work in decomposition (synthesis) of arbitrary permutation unitaries focused on implementation of fully specified permutations defined on all $2^n$ basis states of $n$ qubits, and are exponentially hard to implement in the worst case \cite{shende2003synthesis, soeken2019compiling, khandelwal2024classification}. 

However, when working with sparse quantum states \cite{gonzales2024arbitrary, gleinig2021efficient, luo2024circuit, de2022double, feniou2024sparse, ramacciotti2024simple, mao2024toward, rofougaran2025encoding, mozafari2022efficient}, many of the amplitudes are equal to zero and it does not matter how exactly these zero-amplitudes are permuted among themselves. 
Thus we can implement whatever permutation is easier to implement (e.g. requiring the fewest CX gates, minimizing number of hardware-native gates needed to implement, etc.) as long as it permutes non-zero amplitudes the way we want, which allows us to find more efficient decompositions compared to the case of fully specified permutations.
The algorithm for implementing such permutations in quantum circuits is discussed in Section~\ref{sec:methods}.

One application of sparse state permutation gates is as a building block in a sparse state preparation algorithm, which will be considered in Section~\ref{sec:results}. 

\section{Decomposition Algorithm}
\label{sec:methods}

\begin{figure}
    \centering
    \includegraphics[width=\linewidth]{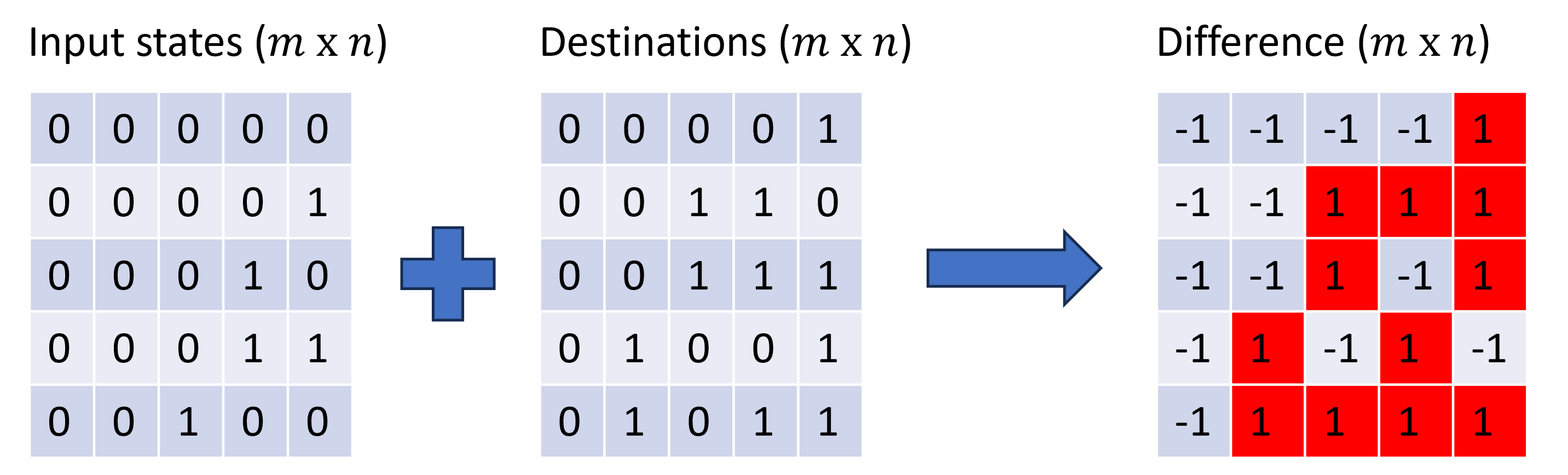}
    \caption{An example of input state labels, their destinations and the corresponding difference matrix.}
    \label{fig:difference_example}
\end{figure}

As an input the algorithm receives description of the desired sparse amplitude permutation, which consists of $m$ amplitude labels and the labels of their corresponding destinations (the two $m$ x $n$ input matrices in Figure~\ref{fig:difference_example}). 
Each label is a bitstring of length $n$ (number of qubits in the system). 
The mapping between the amplitude labels and their destinations must be bijective (one-to-one).

The goal of the algorithm is to produce a quantum circuit made up of hardware-implementable gates such that application of this circuit to a given quantum state produces an output state where each specified amplitude is permuted to its specified destination. Unspecified amplitudes are assumed to be permutationally invariant and can be permuted arbitrarily among themselves. The algorithm attempts to minimize the number of CX gates in the circuit, but does not guarantee optimality.

The goal is achieved by 
\begin{enumerate}
    \item Decomposing the target amplitude permutation gate into a product of multi-controlled X (MCX) gates.
    \item Applying standard Qiskit transpiler \cite{javadi2024quantum} to further decompose MCX gates into hardware-implementable gates (specifically, single-qubit and CX-gates), and simplify the resulting circuit.
\end{enumerate}
Application of Qiskit transpiler is trivial, thus the problem is reduced to step 1.

Based on the input information about amplitude labels and their destinations, one can compute a \textit{difference matrix}, where element $[i, j]$ is equal to 1 if the permutation gate needs to flip $i$-th input amplitude label in position $j$. We say that there is an error in position $[i, j]$ that needs to be fixed. Otherwise, element $[i, j]$ is equal to -1, meaning that flipping state $i$ in bit $j$ will introduce an error instead of fixing it (see Figure~\ref{fig:difference_example}). Thus, the goal can be equivalently reformulated as finding a shortest sequence of operations (MCX gates) that transforms all elements of the difference matrix to -1. Note that all rows of the difference matrix are labeled by the corresponding non-zero amplitudes.

Before we continue, let us introduce some definitions that will be needed to further describe the decomposition algorithm. 
A complete set of all $2^n$ amplitude labels of $n$ qubits can be represented as a hypercube graph, where two labels are connected by an edge if their Hamming distance is equal to 1 (see Figure~\ref{fig:permutation_examples}). 
A \textit{sub-hypercube (SHC)} is a subset of amplitude labels such that these labels also form a hypercube graph. 
Any valid SHC can be compactly described as a subset of \textit{all} amplitude labels conforming to a given \textit{pattern string}, which is a string of $n$ characters that can be either 0, 1 or * (wildcard).
An amplitude label is said to conform to a pattern string if their corresponding characters are equal in all non-wildcard positions.
For example, **1 pattern describes an SHC made up of \{001, 011, 101, 111\} (all bitstrings conforming to the pattern). 
We say that an SHC \textit{spans} the dimensions corresponding to * in its pattern string and is \textit{fixed} in non-wildcard dimensions.

Now let us consider some examples of what kinds of amplitude swaps we can implement in a single \textit{block}, i.e. a sequence of gates consisting of 1 MCX gate optionally conjugated by CX gates. 

\begin{figure*}
    \centering
    \begin{tabular}{C{0.2\textwidth} C{0.24\textwidth} C{0.24\textwidth} C{0.24\textwidth}}
         \includegraphics[height=3cm]{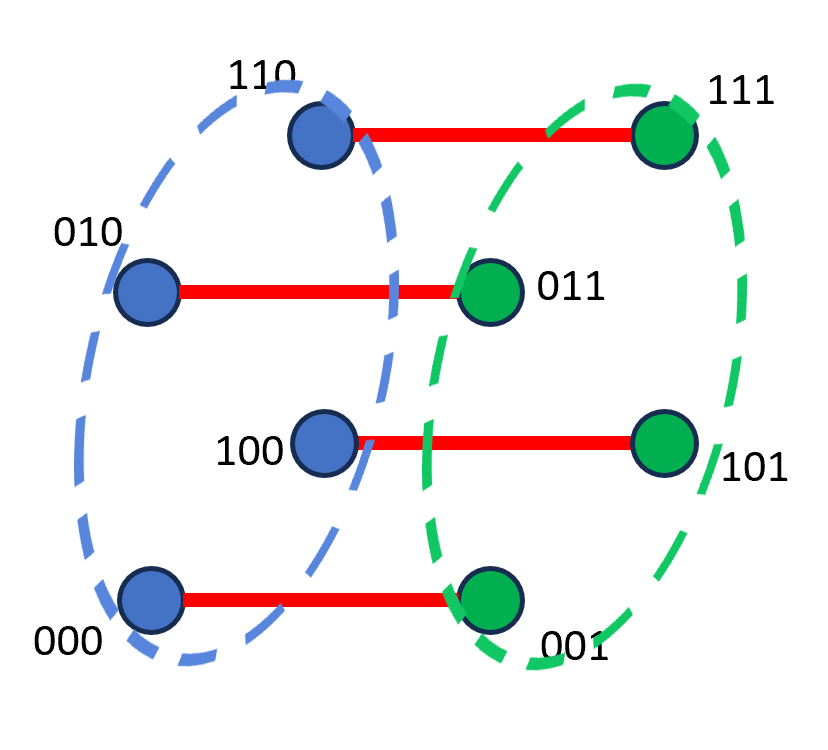} &
         \includegraphics[height=3cm]{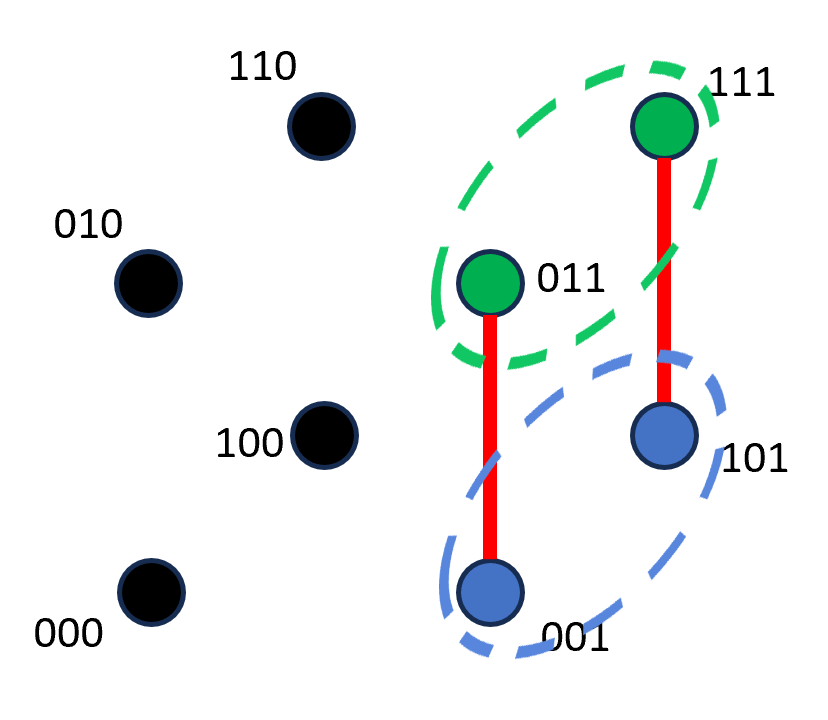} &
         \includegraphics[height=3cm]{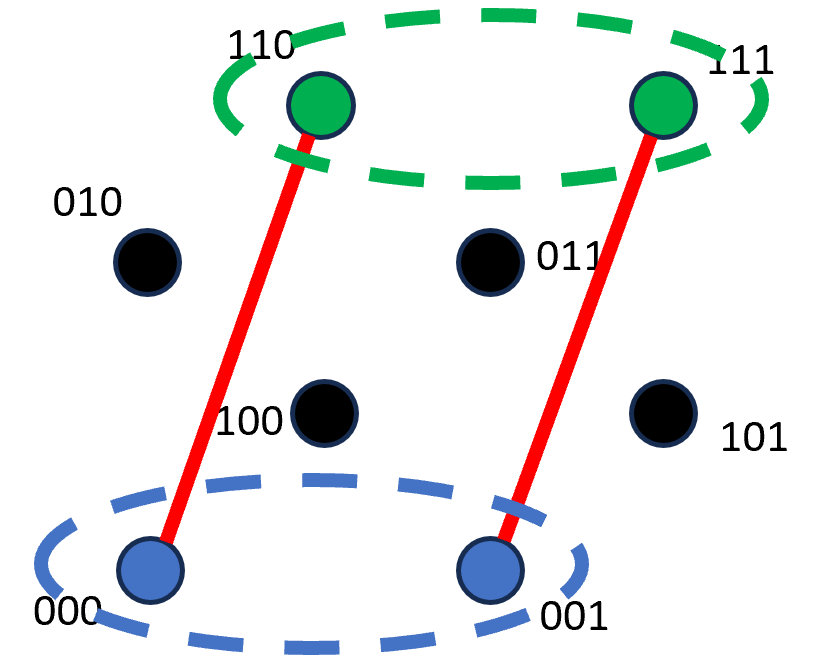} &
         \includegraphics[height=3cm]{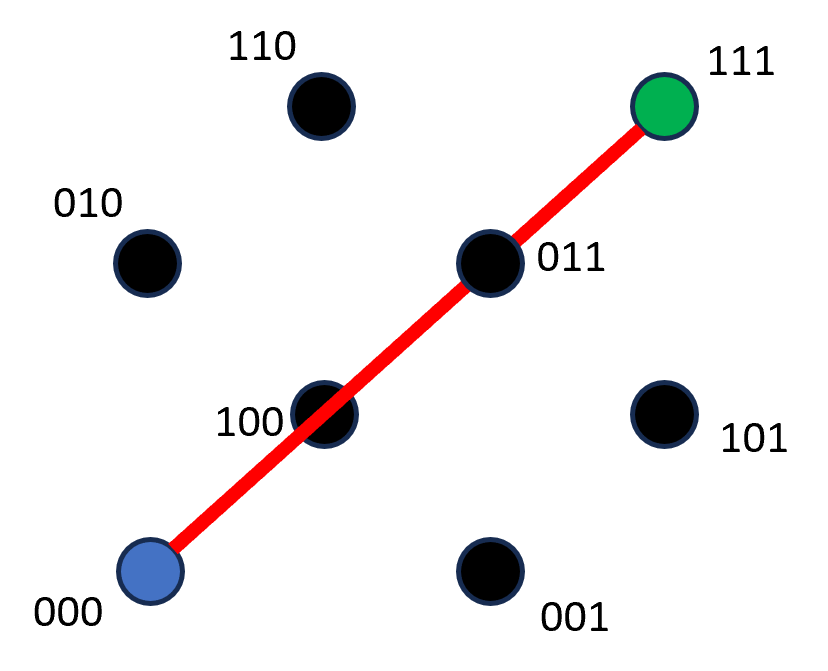} \\
         \includegraphics[height=2cm]{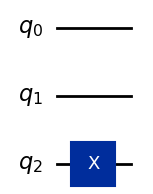} &
         \includegraphics[height=2cm]{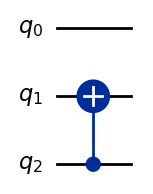} &
         \includegraphics[height=2cm]{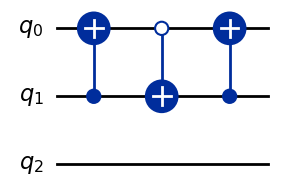} &
         \includegraphics[height=2cm]{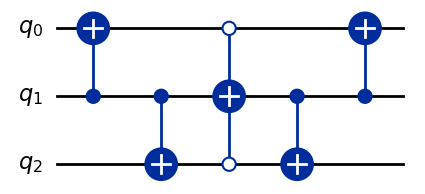} \\
         (a) An uncontrolled X gate permutes the two halves of the hypercube separated by the dimension corresponding to the target qubit. &
         (b) A controlled X gate restricts the permutation to the SHCs corresponding to the state of controls. &
         (c) Conjugating the main MCX gate with a series of CX gates allows us to tilt the swap plane. &
         (d) In the limit of using $n - 1$ controls the corresponding SHCs reduce to single amplitudes and allow us to perform pairwise swaps. \\
    \end{tabular}
    \caption{Several examples of swaps that can be performed on 3 qubits and the quantum circuits implementing them. The two SHCs being swapped are marked with blue and green colors. Red lines show element-wise swaps between individual amplitudes of each SHC. 
    }
    \label{fig:swap_examples}
\end{figure*}

Applying a single uncontrolled X gate to qubit $i$ swaps the two largest SHCs separated by dimension $i$ in the hypercube (i.e. SHCs having wildcards in all positions except $i$). For example, in Figure~\ref{fig:swap_examples}a, placing an X gate on qubit 2 swaps **0 and **1.

Adding controls to an X gate replaces wildcards with the values of controls (0 or 1) in the corresponding positions, thus reducing the size of the swapped SHCs. For example, in Figure~\ref{fig:swap_examples}b, there is a CX gate with the target on qubit 1 and a 1-control on qubit 2, which swaps *01 and *11.

Conjugating a given MCX gate with a sequence of CX gates allows us to tilt the swap plane and flip multiple positions in the affected amplitude labels. Specifically, all gates from the CX gate sequence need to have the 1-control on the target qubit of the main MCX gate and targets on all qubits corresponding to other positions that we want to flip. For example, in Figure~\ref{fig:swap_examples}c, qubit 1 was chosen as the target for the main MCX gate and qubit 0 was added to it by the conjugating CX gates, thus swapping SHCs 00* and 11*.

A more general example is shown in Figure~\ref{fig:swap_examples}d, where we use 2 0-controls on the MCX gate to reduce the size of swapped SHCs to single amplitudes and couple the target qubit with the 2 other qubits, allowing us to swap SHCs 000 and 111.

In general, an arbitrary single-block swap can be described as swapping 2 arbitrary SHCs with the same spanned dimensions. In terms of the difference matrix, it can be described as choosing 2 pattern strings with wildcards in the same positions, and then flipping the signs of all elements at the intersection of rows whose labels conform to one of the pattern strings and columns where the 2 pattern strings are different. Note that flipping the sign of element $[i, j]$ also flips the label of row $i$ in position $j$, which will affect which SHCs can select that row in the subsequent swaps. 


There is a total of $3^n$ SHCs on $n$ qubits (since there is a bijective mapping between the set of all SHCs and the set of all pattern strings), so we cannot explicitly consider every candidate SHC.
Instead, we greedily explore the space of SHCs, starting from the full hypercube that spans all dimensions (*** in case of 3 qubits). 

During the first iteration of the decomposition algorithm we consider all $n$ ways to fix one of the dimensions. 
Each choice generates 2 SHCs corresponding to the value of the fixed dimension, and we pick the pair whose swap will fix the largest number of errors in the difference matrix.

During the second iteration we consider all $n - 1$ ways to fix another dimension, in addition to the dimension chosen in the previous iteration. 
Each choice generates 4 SHCs corresponding to all combinations of values in the 2 fixed dimensions. 
For each SHC and each of its fixed coordinates, we calculate the net number of errors that will be fixed if we flip the value of that coordinate (i.e. column-sum of the difference matrix for rows selected by that SHC).

Then we use that information to build a directed graph where the nodes are SHCs, the edges correspond to particular changes in their fixed coordinates and are initially weighted by the net number of errors that will be fixed if that change happens (\textit{forward weight}).
For each edge SHC1 $\rightarrow$ SHC2, if the target (SHC2) is non-empty (i.e. there are rows in the difference matrix selected by it) then the weight of the backward edge (SHC2 $\rightarrow$ SHC1, \textit{backward weight}) is added to the weight of the forward edge, since we will have to swap them to make it happen. 
This total number of errors is then divided by the number of CX gates required to implement the corresponding swap and used as the final weight of each edge. 

Since we cannot consider all $O(2^n)$ edges from a given node, we apply a similar greedy strategy, where we sort the dimensions by the net number of errors that will be fixed if we flip the corresponding coordinate (in descending order) and then we add one edge for each prefix set of dimensions as long as doing so increases the net number of errors fixed (but at least 1 edge is always added). 
Thus, no more than $n$ edges is added for each SHC node.
Note that each added edge fixes more errors than the previous edge; however, the number of CX gates in their implementation is also larger since they correspond to changes in multiple dimensions (e.g. see Figure~\ref{fig:swap_examples}d).
Therefore the last added edge will not necessarily be the best.

\begin{figure}
    \centering
    \includegraphics[height=3cm]{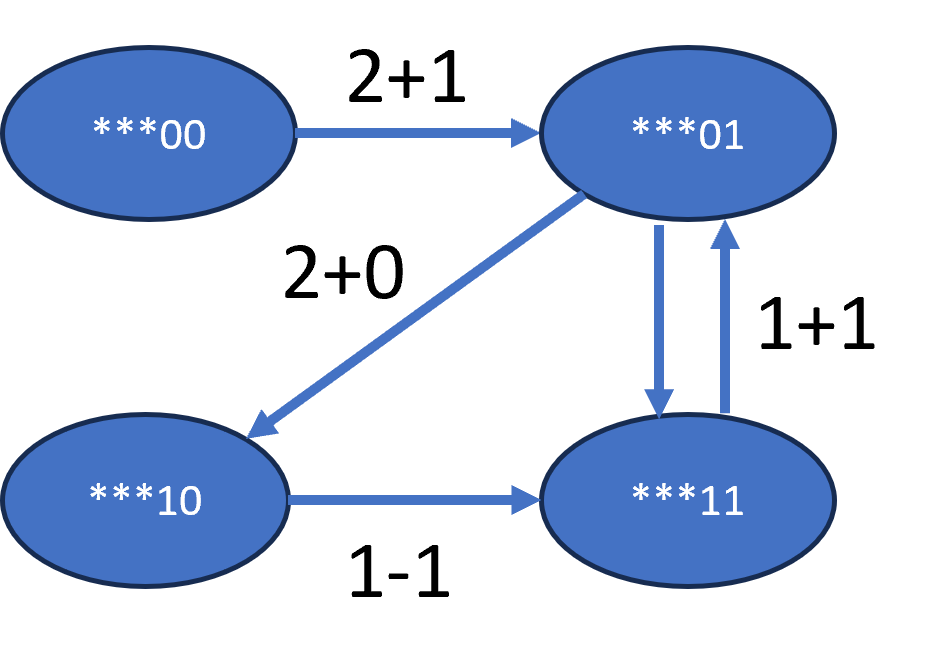}
    \caption{An example of an SHC graph.}
    \label{fig:shc_graph}
\end{figure}

An example of such graph for the input states and difference matrix from Figure~\ref{fig:difference_example} is shown in Figure~\ref{fig:shc_graph}, assuming dimensions 4 and 5 have been chosen as fixed.
The weight of each edge is displayed in the format ``forward weight + backward weight" and is \textit{not} divided by the number of CX gates in this example for simplicity.

SHC ***00 includes 2 states, both of which require a flip in position 2 (counting fixed dimensions only), so we add an edge to ***01. 
The weight of that edge is given by the number of errors that will be fixed by moving ***00 to ***01 (2) and is additionally modified by the number of errors that will be fixed by moving ***01 to ***00 (1), since we cannot move one SHC without moving the other. 
One of the states from ***00 also requires a flip in position 1, but the other one does not, so implementing such flip would result in 0 net errors fixed and the corresponding edge is not added.

SHC ***01 includes 1 state, which requires a flip in both fixed positions. In accordance with our earlier prescription, we add 1 edge for each prefix set of sorted dimensions for as long as doing so increases the net number of errors fixed, which, in this case, results in 1 edge to ***11 (weight 1+1) and 1 more edge to ***10 (weight 2+0).

SHC ***10 includes 1 state, which requires a flip in position 2, so we add an edge to SHC ***11. This time, the state selected by SHC ***11 does not require a flip in position 2, so the forward weight is modified by -1 and results in 0 net weight.

Building these SHC graphs for all $n - 1$ choices of the fixed dimensions that we considered during the second iteration, we select the edge with the largest total weight across all graphs and choose the set of the fixed dimensions corresponding to it as a starting point for the next iteration, where we will consider all $n - 2$ choices to fix the third dimension, etc.

Once we perform all $n$ iterations, we select the best edge found across these iterations, implement the corresponding swap, update the difference matrix and repeat the process until all entries of the difference matrix are equal to -1.
In cases when multiple edges have the same weight, we use forward-only weight and the average number of non-zero adjacent amplitudes in the source SHC as a tiebreakers that help us find better swaps in the subsequent iterations.

Intuitively, forward-only weight works because the negative backward weight is likely to become less negative during the subsequent iterations, as we start considering smaller SHCs.
The average number of non-zero adjacent (Hamming distance = 1) amplitudes is calculated for each non-zero amplitude within SHC and serves as a measure of degree of clustering of non-zero amplitudes within the SHC.
Highly clustered amplitudes are more likely to be selectable together, which will allow us to find good swaps (i.e. swaps corresponding to high-weight edges)
if they all need to be permuted in the same set of dimensions.

Each swap found by this procedure is guaranteed to not increase the total number of errors in the difference matrix. In the worst case the total number of errors can remain the same, but at least 1 amplitude label will be moved to its final destination, therefore the process is guaranteed to finish.

Note that only non-empty SHCs are considered as initial nodes at each iteration, therefore their number does not exceed $m$. 
Empty SHCs can be added to the graph if there is an edge pointing to it, but the total number of edges from each node in an SHC graph is also limited.

The algorithm requires no more than $O(n+m)$ swaps, each of which needs $O(n)$ iterations. 
Each iteration considers $O(n)$ choices of a new dimension to fix, and for each choice we build an SHC graph with $O(m)$ initial nodes and $O(nm)$ edges. 
For each node we sort the errors across its fixed coordinates in $O(n\log(n))$. 
Thus, the overall classical complexity of the algorithm is $O((n+m)n^3m\log(n))$.

Each of the swaps can be decomposed with $O(n)$ CX gates using 1 ancilla qubit, therefore the generated permutation circuit requires $O(n)$ qubits and its depth and CX count are $O((n + m)n)$. 

Pseudocode of the decomposition algorithm is shown in Algorithm~\ref{alg:decomposition}.

\begin{algorithm}
    \DontPrintSemicolon
    \caption{Decomposition of Sparse Permutation Gates}
    \label{alg:decomposition}
    \KwIn{$a \gets m$ x $n$ matrix of amplitude labels \newline
    $b \gets m$ x $n$ matrix of amplitude destinations}
    \KwOut{$c \gets $Quantum circuit implementing the permutation specified by $a$ and $b$}
    $c \gets $ empty circuit \;
    $d \gets CalculateDifferenceMatrix(a, b)$ \;
    \While{\textnormal{any element of} d = 1} {
        $fixed \gets$ [] \;
        $swaps \gets$ [] \;
        \For{$numFixed \gets 1...n$} {
            $swaps \gets swaps$ $\oplus$ None \;
            \For{$i \gets Set(\textnormal{1...n}) - Set(fixed)$} {
                $newFixed \gets fixed$ $\oplus$ $i$ \;
                $g \gets FormSHCGraph(newFixed, d)$ \;
                \If{\textnormal{$BestEdge(g) > Last(swaps)$}} {
                    $Last(swaps) \gets BestEdge(g)$
                }
            }
            $fixed \gets FixedDimensions(Last(swaps))$ \;
        }
        $bestSwap \gets Best(swaps)$ \;
        $c \gets ImplementSwap(bestSwap, c)$ \;
        $d \gets UpdateDifferenceMatrix(bestSwap, d)$ \;
    }
\end{algorithm}

\section{Application: Preparation of Sparse Clustered Quantum States}
\label{sec:results}

One application of the amplitude permutation gate is as a building block in a sparse quantum state preparation algorithm. Specifically, an arbitrary sparse state can be prepared as follows:
\begin{enumerate}
    \item Find a permutation that permutes the target sparse state into a locally dense state within some SHC.
    \item Use any dense state preparation method on a subset of qubits corresponding to the found SHC to prepare the permuted version of the state.
    \item Apply inverse permutation to permute the SHC into the target sparse state.
\end{enumerate}

\begin{figure}
    \centering
    
    \subfloat[Cluster-preserving permutation.]{\includegraphics[height=3cm]{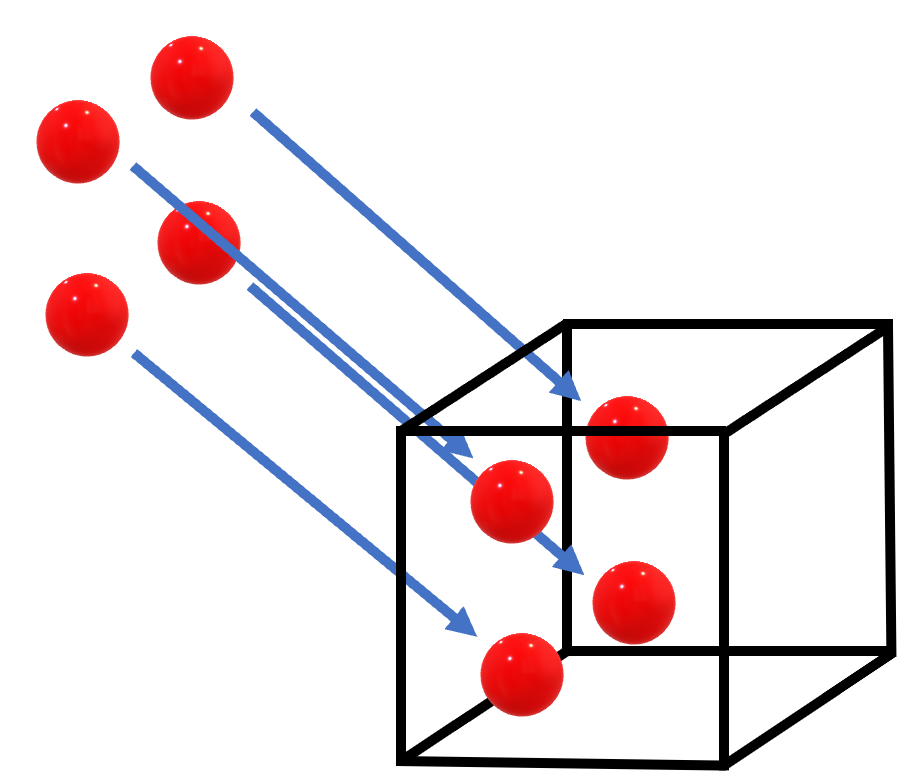}}
    \hfill
    \subfloat[Arbitrary permutation.]{\includegraphics[height=3cm]{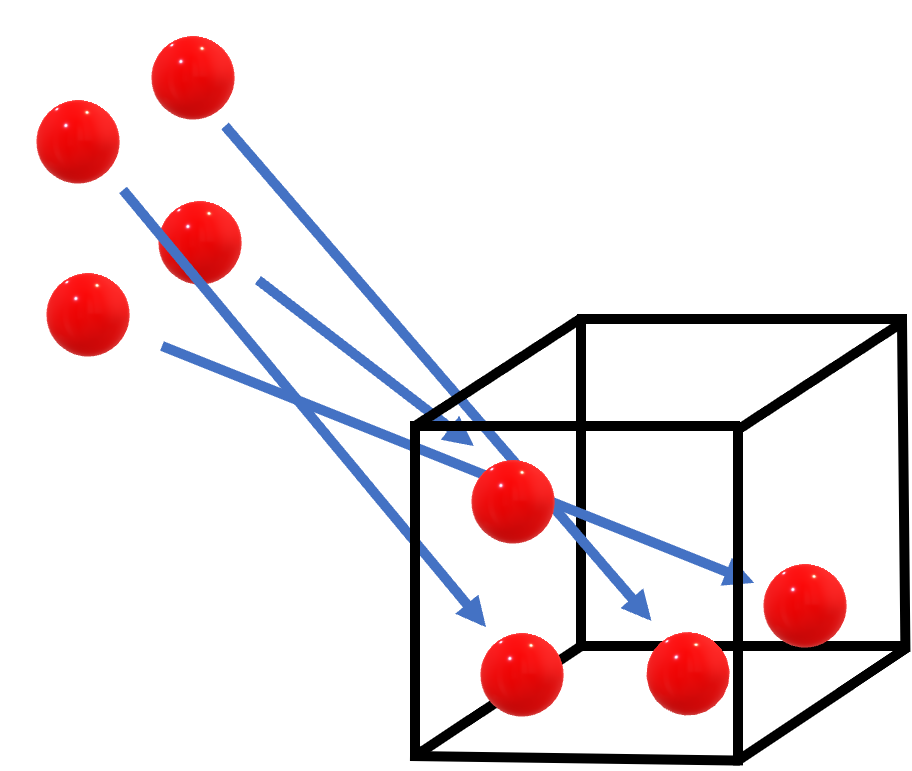}}
    
    \caption{Examples of permutations that move a cluster of non-zero amplitudes (red spheres) into a locally dense SHC (black cube).}
    \label{fig:cluster_permutation_examples}
\end{figure}

The algorithm presented in Section~\ref{sec:methods} implements an arbitrary given permutation.
However, the only requirement for a valid permutation in step 1 is to move all non-zero amplitudes into a locally dense SHC, does not matter in what order.
Therefore there are exponentially many valid permutations that can be used in step 1 with different implementation complexity that depends on the structure of the permutation.
Namely, if there is a local \textit{cluster} of non-zero amplitudes then it is easier to move that cluster as a whole, without permuting it internally, which will reduce both the total number of swaps and the number of controls on MCX gates. 
Therefore, an efficient permutation for step 1 has to preserve cluster structure, and we need to design an algorithm that can find such permutations (see Figure~\ref{fig:cluster_permutation_examples}). 

\subsection{An Algorithm for Finding Cluster-Preserving Permutations}
\label{sec:cluster_preserving_algorithm}

There is an exponential number of possible permutations of $m$ non-zero amplitudes into a locally dense SHC.
Therefore, similarly to the gate decomposition algorithm, we are going to use a greedy approach, which does not guarantee optimality, but finds a reasonably good solution.
First, we calculate $n' = \lceil\log_2(m)\rceil$, the number of spanned dimensions (qubits) in the smallest SHC necessary to contain all $m$ non-zero amplitudes. 
Then we find an SHC spanned by $n'$ qubits that contains the largest number of non-zero amplitudes.

Specifically, we start from the full $n$-dimensional hypercube and, at each of the next $n - n'$ iterations, greedily choose the next dimension to fix as the one that leaves the largest number of non-zero amplitudes in the remaining SHC.
If multiple candidates have the same number of non-zero amplitudes, then we use the average number of adjacent non-zero amplitudes as a tiebreaker.

This $n'$-dimensional SHC is added as the first element to the list of currently existing SHCs.
Then, for as long as there are non-zero amplitudes not covered by any of the SHCs from the current list, we take the most sparse SHC ($S$) from the list (i.e. the SHC with the largest difference between its full capacity and the number of non-zero amplitudes currently covered by it), and split it into 2 children SHCs along one of the dimensions spanned by $S$ (the children replace $S$ in the list of SHCs).
The splitting dimension is chosen greedily to maximize the number of non-zero amplitudes covered by the first child of $S$.
If the second child is empty (i.e. all amplitudes of $S$ are covered by its first child), then we change its fixed coordinates to a new set chosen to cover the largest number of still-uncovered non-zero amplitudes.

Throughout this process we keep track of the parent-children tree and, once all non-zero amplitudes are covered by SHCs, we use it for backtracking and find a cluster-preserving permutation of the target sparse state into the original $n'$-dimensional SHC.

Finding the initial $n'$-dimensional SHC requires $O(n)$ iterations with $O(nm)$ operations on each iteration to count the number of non-zero amplitudes in each candidate SHC.
After that, building a splitting tree of SHCs covering all non-zero amplitudes requires $O(m)$ splits with $O(\log(m))$ overhead to find the most sparse SHC.
For each split we need $O(nm)$ operations to choose the best splitting dimension and move the second child.
The final permutation can be recovered from the splitting tree in $O(nm)$.
Thus, overall complexity of the algorithm is $O(n^2m+nm^2\log(m))$.

Pseudocode of the cluster-preserving permutations algorithm is shown in Algorithm~\ref{alg:cluster-preserving}.

\begin{algorithm}
    \DontPrintSemicolon
    \caption{Cluster-Preserving Permutations}
    \label{alg:cluster-preserving}
    \KwIn{$b \gets m$ x $n$ matrix of amplitude labels in the target state}
    \KwOut{$a \gets m$ x $n$ matrix of corresponding amplitude labels in a dense SHC}
    $n' \gets \lceil\log_2(m)\rceil$ \;
    $shc \gets $ * repeated $n$ times \;
    \For{$i \gets 1...(n-n')$} {
        $nextSHCs \gets []$ \;
        \For{$j \gets SpannedDims(shc)$} {
            $nextSHCs \gets nextSHCs \oplus Fix(shc,\ j,\ 0)$ \;
            $nextSHCs \gets nextSHCs \oplus Fix(shc,\ j,\ 1)$
        }
        $shc \gets Best(nextSHCs)$ \;
    }
    $shcs \gets shc$ \;
    \While{\textnormal{not $IsCovered(b,\ shcs)$}} {
        $S \gets MostSparse(shcs)$ \;
        $Remove(S,\ shcs)$ \;
        $splittings \gets $ None \;
        \For{$i \gets SpannedDimensions(S)$} {
            $splittings \gets splittings \oplus Split(S,\ i)$
        }
        $child1,\ child2 \gets Best(splittings)$ \;
        \If{Empty(child2)} {
            $child2 \gets Move(child2,\ b)$
        }
        $shcs \gets shcs \oplus child1 \oplus child2$ \;
    }
    $a \gets RecoverPermutation(shcs)$
\end{algorithm}

\begin{figure}
    \centering
    \includegraphics[height=4cm]{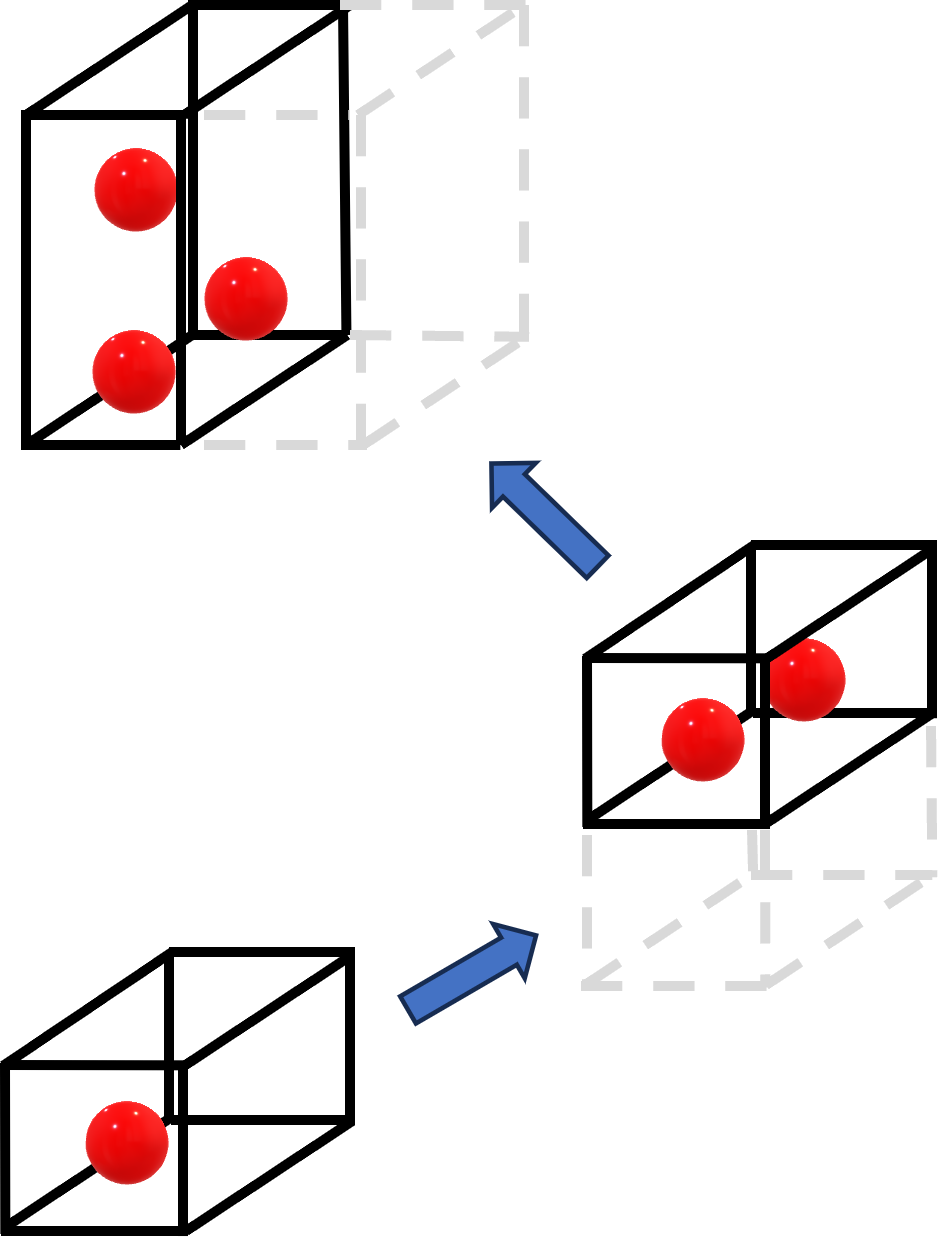}
    \caption{An example of finding cluster-preserving permutations.}
    \label{fig:permutation_algorithm_example}
\end{figure}

An example of algorithm is shown in Figure~\ref{fig:permutation_algorithm_example}. The initial 3-dimensional SHC covered 3 non-zero amplitudes (red spheres). Only the left half of it is sufficient to cover these amplitudes, so the right half was moved to cover cluster of 2 amplitudes. Then the same process was applied again to the right half and its bottom half was moved to cover the last amplitude. Backtracking these splits, we can find a cluster-preserving permutation of the non-zero amplitudes into the original SHC. In this case, the largest cluster does not need to be moved anywhere since the dense SHC was built around it.

\subsection{Numerical Benchmarking}

\begin{figure}
    \centering
    \includegraphics[width=\linewidth]{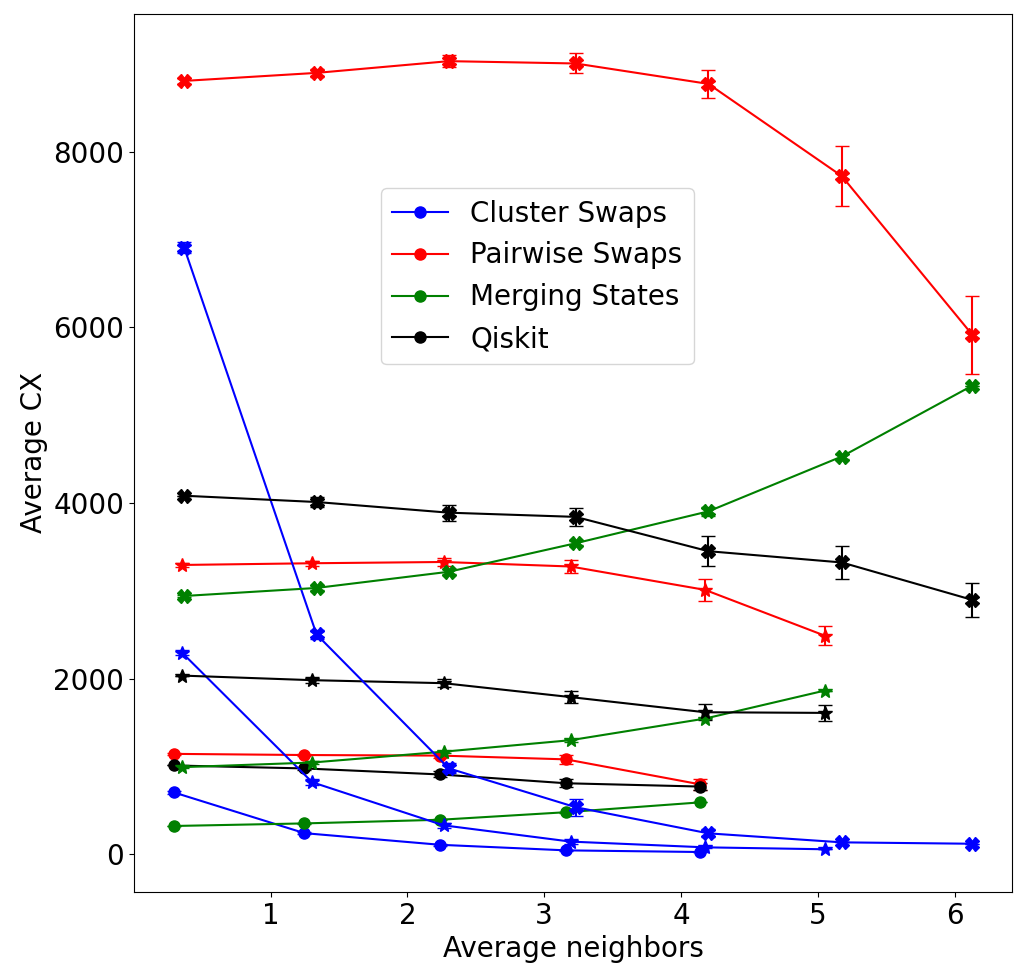}
    \caption{Average number of CX gates necessary to prepare random sparse states with varying degree of clustering for different state preparation methods. Circles/stars/crosses mark datasets where the target states are generated on 10/11/12 qubits and have $2^5$/$2^6$/$2^7$ non-zero amplitudes (5/6/7 datasets total).}
    \label{fig:benchmarking}
\end{figure}

The described approach was used to prepare multiple sets of random sparse states with various degree of clustering. 
The degree of clustering of a given state is measured as the number of adjacent (Hamming distance = 1) non-zero amplitudes averaged across all of its non-zero amplitudes.
In Figure~\ref{fig:benchmarking}, we show the average number of CX gates necessary to prepare random states from a given dataset as a function of the average number of neighbors (degree of clustering) in that dataset.
Each type of marker and the value of average neighbors corresponds to a separate dataset consisting of 100 random sparse states (18 datasets total).
Error bars show 95\% confidence intervals for the y-values.

The following methods have been considered:
\begin{enumerate}
    \item Cluster Swaps. 
    This is the method described in this section that finds a cluster-preserving permutation and uses it as an input for the MCX decomposition algorithm for amplitude permutation gates described in Section~\ref{sec:methods}. 
    Qiskit's \textit{prepare\_state} was used as a dense state preparation method in step 2 of the algorithm.
    \item Pairwise Swaps. 
    An alternative decomposition scheme for an arbitrary unitary is to decompose it into a product of 2-level unitaries \cite{nielsen2010quantum}. 
    In terms of permutations, a 2-level unitary corresponds to a pairwise swap, so this method shows the performance in case if we perform pairwise swaps only.
    \item Merging States. 
    Alternatively, we might decide to forgo permutations entirely and just use a sparse state preparation algorithm to prepare the target states directly. 
    Specifically, we chose a state-of-the-art sparse state preparation algorithm called Merging States \cite{gleinig2021efficient} as a reference in this category.
    \item Qiskit.
    Another alternative is to apply a dense state preparation algorithm to all qubits (no permutations). 
    Specifically, we used the same \textit{prepare\_state} method from Qiskit library \cite{qiskit2024} as for Cluster Swaps method.
    Note that this is the only method (among the methods considered here) most sensitive to the total number of qubits rather than number of non-zero amplitudes.
\end{enumerate}

As one can see from Figure~\ref{fig:benchmarking}, the efficiency of all methods, except Merging States, improves as the degree of clustering is increased, but Cluster Swaps method of this paper is the most sensitive to it.

In the left-hand side of Figure~\ref{fig:benchmarking}, we can see that for the states where the average number of neighbors is $<$ 1 Cluster Swaps method is less efficient than Merging States.
These are the states with no cluster structure, i.e. the states where majority of non-zero amplitudes have no other adjacent non-zero amplitudes.

However, for the states with at least minimal cluster structure, where the average number of neighbors $\geq$ 1, Cluster Swaps method is able to outperform all other alternatives considered here.
An example of such a state is a state where all non-zero amplitudes come in pairs, i.e. for each non-zero amplitude there is 1 other adjacent non-zero amplitude. 
Additionally, the separation between Cluster Swaps and other methods appears to increase as a function of the total number of qubits (different symbols), which suggests that the Cluster Swaps method will retain its advantage asymptotically as well.

\section{Conclusion}
\label{sec:conclusion}

In this paper we considered a novel method for decomposition of sparse amplitude permutation gates, which can serve as useful primitives for quantum computing algorithms.
As an example of application, we have considered applying it in the design of a new sparse state preparation algorithm, Cluster Swaps, which is able to efficiently prepare quantum states, where non-zero amplitudes tend to be adjacent to one another, thus forming multiple scattered clusters of non-zero amplitudes.

Preparation of such states can be useful for amplitude encoding of data \cite{abohashima2020classification, ranga2024quantum, pande2024comprehensive, wu2025quantum} in quantum machine learning  applications where data vectors are typically high-dimensional and sparse \cite{fan2008liblinear, araki2017accelerating, poulinakis2023machine} (e.g. text classification \cite{agarwal2014text, el2013understanding}, recommender systems \cite{cheng2016wide}, computer vision \cite{alskeini2018face}, bioinformatics \cite{xiao2018semi}), and could be clustered depending on specific problem instances and arrangement of data vector elements. 

Future work on this algorithm can include finding better heuristics to decompose the target permutation gate.
For example, instead of just selecting a single best edge among the SHC graphs, one could take into account higher-order patterns, such as chains with an empty SHC at the end or cycles (see Figure~\ref{fig:shc_graph}), which would allow to avoid the cost of the backward edges and potentially find better swaps.

\section*{Author contributions}
I.G.: Study design, Code, Data acquisition, Figures, Original draft of the text

R.H.: Supervision, Text review and editing, Funding acquisition

All authors wrote, read, edited, and approved the final manuscript.

\section*{Acknowledgments}
R. Herrman acknowledges DE-SC0024290. The funder played no role in study design, data collection, analysis and interpretation of data, or the writing of this manuscript. 

\section*{Competing interests}
All authors declare no financial or non-financial competing interests. 

\section*{Code and Data Availability}
The code and data for this research can be found at \\ 
\small{\url{https://github.com/GaidaiIgor/permutation_unitaries}}
\clearpage
\bibliographystyle{unsrt}
\bibliography{refs}

\end{document}